\documentclass[aps,prd,reprint,preprintnumbers,floats,epsfig,nofootinbib,amssymb,
nofootinbib]{revtex4}

\usepackage{slashed}
\usepackage{graphicx,color}
\usepackage{epsfig}
\usepackage{subfigure}
\usepackage{epsfig}
\usepackage{epstopdf}
\usepackage{dcolumn}
\usepackage{bm}
\usepackage{color}
\usepackage{ulem}
\usepackage{multirow}
\usepackage{enumitem}
\usepackage[colorlinks,
            linkcolor=black,
            anchorcolor=black,
            citecolor=black
            ]{hyperref}

\begin{document}

\baselineskip=15pt

\title{Muon collider  signatures for a $Z'$ with a maximal  $\mu - \tau$ coupling in $U(1)_{L_\mu - L_\tau}$}

\author{Jin Sun$^{1,3}$\footnote{019072910096@sjtu.edu.cn}}
\author{Fei Huang${}^{4,1}$\footnote{fhuang@sjtu.edu.cn}}
\author{Xiao-Gang He${}^{1,2}$\footnote{hexg@sjtu.edu.cn}}

\affiliation{${}^{1}$Tsung-Dao Lee Institute, and School of Physics and Astronomy, Shanghai Jiao Tong University, Shanghai 200240, China}
\affiliation{${}^{2}$National Center for Theoretical Sciences, and Department of Physics, National Taiwan University, Taipei 10617, Taiwan}
\affiliation{${}^{3}$Center for Theoretical Physics of the Universe, Institute for Basic Science, Daejeon 34126, Korea}
\affiliation{${}^{4}$School of Physics and Technology, University of Jinan, Jinan, Shandong 250022, China}

\begin{abstract}
The gauged $U(1)_{L_\mu - L_\tau}$ model is a candidate model for explaining the muon g-2 anomaly because the $Z'$ in the model has a natural normal coupling to muon. Due to other experimental data constraints the viable mass range for the usual $Z'$ is constrained to be lower than a few hundred MeV. It has been shown that if the $Z'$ has a maximal off-diagonal mixing,
$(\bar \mu \gamma^\mu \tau + \bar \tau \gamma^\mu \mu) Z'_\mu$, a large mass for $Z'$ is possible. This class of models has a very interesting signature for detection, such as $\mu^-\mu^+ \to \tau^- \tau^+$ pair, $\mu^- \mu^+ \to \mu^\pm\mu^\pm \tau^\mp \tau^\mp$ at a muon collider. In this work we study in detail  these processes. We find that the in the parameter space solving the muon g-2 anomaly, t-channel $\tau^- \tau^+$ pair production  can easily be distinguished at more than 5$\sigma$  level from the s-channel production as that predicted in the standard model. The smoking gun signature of doubly same sign $\mu^\pm \mu^\pm + \tau^\mp\tau^\mp$ pairs production can have a  5$\sigma$ sensitivity, at a muon collider of 3 TeV with $\mathcal{O}$($fb^{-1}$) luminosity.
\end{abstract}

\maketitle

\section{Introduction}
 
Muon collider with a multi TeV energy and high luminosity of order $\mathcal{O}$($fb^{-1}$) will provide excellent chances for new particles and new interactions beyond the standard model (SM).  Search for a new massive gauge boson, generically referred to as $Z'$, is one of the interesting topics in this regard. Among many of $Z'$ models, the gauged $L_\mu-L_\tau$ model has received a lot of attention due to the possibility of exchanging the $Z'$ to explain the muon g-2 anomaly~\cite{Muong-2:2021ojo}.  Although this anomaly may be due to theoretical uncertainties from QCD contributions, the $U(1)_{L_\mu-L_\tau}$ $Z'$ can explain this discrepancy~\cite{Baek:2001kca, Ma:2001md, Gninenko:2001hx, Pospelov:2008zw, Heeck:2011wj, Harigaya:2013twa, Altmannshofer:2014pba, Altmannshofer:2016oaq} in certain regions of parameter space. In this paper we study how a muon collider can provide useful information for $Z'$ in one of such models from gauge $U(1)_{L_\mu - L_{\tau}}$.

It has been shown that in order to explain the muon $(g-2)_\mu$,  a strong constraint on the $L_\mu-L_\tau$ comes from the neutrino trident process $v_\mu +N\to v_\mu +N +\mu^+ \mu^-$~\cite{CCFR:1991lpl,CHARM-II:1990dvf,NuTeV:1999wlw} with $(\bar \mu \gamma^\mu \mu- \bar \tau \gamma^\mu \tau + \bar \nu^\mu_L \gamma^\mu \nu^\mu_L - \bar \nu^\tau_L \gamma^\mu \nu^\tau_L)Z'_\mu$, which severely constrains the $Z'$ mass $m_{Z'}$ with $m_{Z'}<300$ MeV~\cite{Cen:2021ryk}.
 Recently a new mechanism has been proposed to widen  $Z'$ mass range for  resolving the muon $(g-2)_\mu$ anomaly~\cite{Cheng:2021okr}, which transforms the original flavor-conserving $Z'$ interaction into the fully off-diagonal one, 
$(\bar \mu \gamma^\mu \tau+ \bar \tau \gamma^\mu \mu + \bar \nu^\mu_L \gamma^\mu \nu^\tau_L + \bar \nu^\tau_L \gamma^\mu \nu^\mu_L)Z'_\mu$,
 to evade constraint from neutrino trident process and therefore to lift the upper bound for $Z'$ mass.  In this case, $Z'$ exchange can induce $\tau \to \mu \bar \nu \nu$ with a larger branching ratio from experimental data. This conflict can be resolved by introducing  type-II seesaw $SU(2)_L $ triplet scalars. 
Because of the model's special coupling with muon, there are very distinctive signatures at a muon collider, such as the t-channel production of $\mu^-\mu^+ \to \tau^- \tau^+$ while in the SM it is an s-channel process,  and  a smoking gun signature of doubly same sign muon and tau pairs production, $\mu^- \mu^+ \to \mu^\pm\mu^\pm \tau^\mp \tau^\mp$.
In the following sections, we provide our findings in carry detail.

\section{The $U(1)_{L_{\mu}-L_{\tau}}$ model for maximal $\mu$-$\tau$ coupling}

In the simplest $U(1)_{L_\mu - L_\tau}$ model, the left-handed $SU(3)_C\times SU(2)_L\times U(1)_Y$ doublets $L_{L\;i}: (1, 2, -1/2)$ and the right-handed singlets $e_{R\;i}: (1,1,-1)$ transform under the gauged $U(1)_{L_\mu-L_\tau}$ group as $0,\;1,\;-1$ for the first, second and third generations, respectively. The $Z^\prime$ gauge boson of the model only interacts with leptons in the  weak interaction basis~\cite{lmu-ltau1, lmu-ltau2}
\begin{eqnarray}
{\cal L}_{Z'}=- \tilde g (\bar \mu \gamma^\mu \mu - \bar \tau  \gamma^\mu \tau + \bar \nu_\mu \gamma^\mu L \nu_\mu - \bar \nu_\tau \gamma^\mu L \nu_\tau) Z^\prime_\mu \;, \label{zprime-current}
\end{eqnarray}
where $\tilde g$ is the $U(1)_{L_\mu - L_\tau}$ gauge coupling, and $L(R) = (1 - (+)\gamma_5)/2$.
Introducing a scalar $S$ transforming as a singlet under the SM gauge group, but with $U(1)_{L_\mu - L_\tau}$ charge $1$, after $S$ develops a vacuum expectation value $v_S/\sqrt{2}$, $Z'$ will obtain a mass $m_{Z'} = \tilde g v_S$. 

Exchange $Z'$ at one loop level can generate a non-zero anomalous muon g-2 which can explain the anomaly observed. However, $Z'$
exchange will produce a non-zero contribution to the neutrino trident  process  $v_\mu +N\to v_\mu +N +\mu^+ \mu^-$. Neutrino trident  data then constrain the $Z'$ mass to be less than 300 MeV~\cite{Cen:2021ryk}. 
To avoid neutrino trident  data constraint on the $Z'$ interaction~\cite{Cheng:2021okr}, one introduces new scalar particles to make the $Z'$ interaction to muon and tauon off diagonal so that the neutrino trident  process will not happen at tree level.  Such a model had been proposed a long time ago with the bits of help of three Higgs scalars~\cite{Foot:1994vd}. 

We briefly outline the steps to obtain such a model.  One needs to introduce three Higgs doublets $H_{1,2,3}: (1,2, 1/2)$ ($<H_i> = v_{i}/\sqrt{2}$) with   $U(1)_{L_\mu - L_\tau}$ charges ($0, 2, -2$)
and to impose an unbroken exchange symmetry $Z^\prime \to - Z^\prime$,  $H_1 \leftrightarrow H_1$ and $H_2 \leftrightarrow H_3$ with $v_2=v_3=v$,  to do the job.  In this case the $Z^\prime$ interaction and Yukawa terms to leptons are given by
\begin{eqnarray}
L _{H}= &&- \tilde g (\bar l_2 \gamma^\mu L l_2- \bar l_3  \gamma^\mu L l_3 + \bar e_2 \gamma^\mu R e_2 - \bar e_3 \gamma^\mu R e_3) Z^\prime_\mu\nonumber\\
&&- [Y^l_{11} \bar l_1 R e_1 + Y^l_{22} (\bar l_2 R e_2 +\bar l_3 R e_3 ) ] H_1 
-Y^l_{23} (\bar l_2 R e_3 H_2 +\bar l_3 R e_2 H_3 ) + H.C.
\end{eqnarray}
The transformation between  the charged lepton mass eigen-state and weak eigen-state basis is given by
\begin{eqnarray}\label{eigen}
\left (
\begin{array}{c}
\mu\\
\tau
\end{array}
\right )
= {1\over \sqrt{2}}
\left (
\begin{array}{rr}
1&\;-1\\
1&\;1
\end{array}
\right )
\left (\begin{array}{c}
e_2\\
e_3
\end{array}
\right ),\;
\end{eqnarray}
Similar transformation applies to neutrinos. In the new basis, the $Z^\prime$ interactions with leptons become the following form as desired,
\begin{eqnarray}
{\cal L}_{Z'}=	- \tilde g (\bar \mu \gamma^\mu \tau +  \bar \tau  \gamma^\mu \mu + \bar \nu_\mu \gamma^\mu L \nu_\tau  + \bar \nu_\tau \gamma^\mu L \nu_\mu) Z'_\mu \;. \label{zprime-changing}
\end{eqnarray}

The above $Z'$ interaction will lead to  a much larger $\tau \to \mu \bar \nu_\mu \nu_\tau$ branching ratio, which is excluded by experimental value by more than 5$\sigma$ if the muon g-2 anomaly is explained by $Z'$ exchange at one loop level. Therefore one needs to reduce the branching ratio for this decay to satisfy the experimental constraint while adressing $(g-2)_\mu$ anomaly. 

This problem  can be solved by  introducing three $Y=1$ triplet scalar $\Delta_{1,2,3}: (1,3,1)$ ($<\Delta_i> =  v_{\Delta i}/\sqrt{2}$)  with $U(1)_{L_\mu-L_\tau}$ charges ($0,-2,2$)~\cite{Cheng:2021okr}. 
This $\Delta$ field is the famous  type-II seesaw mechanism providing small neutrino masses~\cite{Lazarides:1980nt,Mohapatra:1980yp,Konetschny:1977bn,Cheng:1980qt,Magg:1980ut,Schechter:1980gr} with 
the component fields 
\begin{eqnarray}
	\Delta = \left (\begin{array}{cc}   \Delta^+/\sqrt{2}&\;  \Delta^{++}\\  \Delta^0&\; - \Delta^+/\sqrt{2} \end{array} \right )\;,\;\;\;\Delta^0=\frac{v_\Delta+\delta+i\eta}{\sqrt{2}}\;. 
\end{eqnarray}
Under the above exchange symmetry $\Delta_1 \leftrightarrow  \Delta_1,\; \Delta_2 \leftrightarrow  \Delta_3$ with $v_{\Delta 2} = v_{\Delta 3}$, the Yukawa terms in the basis shown in Eq.(\ref{eigen}) are 
\begin{eqnarray}
L_\Delta = -&& \left [\bar l^c_\mu L l_\mu (Y^\nu_{22}( \Delta_2 +\Delta_3) - 2 Y^\nu_{23} \Delta_1) 
 + \bar l^c_\tau L  l_\tau (Y^\nu_{22} (\Delta_2 +\Delta_3)  + 2 Y^\nu_{23} \Delta_1)
+ 2\bar l^c_\mu L  l_\tau (Y^\nu_{22}( \Delta_2 -\Delta_3))
\right]/2 + H.C.\;.
\end{eqnarray}
Expanding out the above interaction in terms of the component fields $\Delta^{0, +, ++}$, we obtain  
\begin{eqnarray}
&&
L_\Delta = - (\bar \nu_e^c, \bar \nu_\mu^c, \bar \nu_\tau^c ) M(\Delta^0)  L
\left ( \begin{array}{c}
\nu_e \\ \nu_\mu \\ \nu_\tau
\end{array}
\right )
+ \sqrt{2} (\bar \nu_e^c, \bar \nu_\mu^c, \bar \nu_\tau^c ) M(\Delta^+) L
\left ( \begin{array}{c}
e \\ \mu \\ \tau
\end{array} \right )
+ (\bar e^c, \bar \mu^c, \bar \tau^c ) M(\Delta^{++}) L
\left ( \begin{array}{c}
e \\ \mu \\ \tau
\end{array} \right )
\;,\nonumber\\
&&\mbox{with}\;\;\; M(\Delta)  = \left ( \begin{array}{ccc}
Y^\nu_{11} \Delta_1&0&0\\
\\
0& (Y^\nu_{22}(\Delta_2 +\Delta_3) - 2 Y^\nu_{23}\Delta_1)/2 & Y^\nu_{22}(\Delta_2-\Delta_3)/2\\
\\
0&Y^\nu_{22}(\Delta_2-\Delta_3)/2&(Y^\nu_{22}(\Delta_2 +\Delta_3) + 2 Y^\nu_{23}\Delta_1)/2
\end{array}
\right ) \;.
\end{eqnarray}
Here we find that if assuming the degenerate case $\Delta_2=\Delta_3$ required by the exchange symmetry $\Delta_2 \leftrightarrow  \Delta_3$ and $Y_{11,23}<<Y_{22}$, $M(\Delta)$ is diagonal matrix with non-zero entries $M_{22}\approx M_{33}$. This is helpful for simplifying the our model annalysis.

Therefore, the scalar sectors include three Higgs doublets and three triplet scalars. To simplify the analysis, we will make the following assumptions: $Y_{11,23}<<Y_{22}$, degenerate case $m_{\Delta 2}=m_{\Delta 3}$  and other heavier new degrees of freedom. Under these assumptions, the new scalar effects on SM particles will be dominated by $\Delta_2$ interaction terms.  If further assuming the degenerate triplet components, we can obtain $ m_{\Delta^{++}}= m_{\Delta^+}=  m_{\Delta^0}  
	 = m_\Delta$. In this case, 
 for $\Delta m=0$ with large $v_\Delta \sim O(\mbox{GeV})$, doubly-charged scalar mass below 420 GeV has already excluded from the collider constraints~\cite{Ashanujjaman:2021txz}. Therefore, we will focus on the scenario with $m_\Delta > 420$ GeV.

Note that  our models contain the totally flavor changing $Z'$ interactions and $\Delta$ mediated flavor conserving interactions. The relevant low-energy phenomenology has been studied in Ref.~\cite{Cheng:2021okr}. We will mainly focus on the muon collider aspects in the following part. We firstly focus on the two body case $\mu^-\mu^+ \to \tau^- \tau^+$, then further expand to four body case $\mu^-\mu^+ \to \mu^\pm\mu^\pm + \tau^\mp \tau^\mp$. The influences can be expressed in the four model parameters. The two ones are $\tilde g$, $m_{Z'}$ from $U(1)_{L_\mu-L_\tau}$, and the remaining two  are $Y_{22}$, $m_\Delta$ from triplet scalar. The total contributions for the above processes should contain the  $Z'$ effects by $\tilde g$ and $m_{Z'}$, and triplet ones by  $Y_{22}$ and $m_\Delta$ simultaneously. Therefore, the dominant  contribution could come from the $Z'$ effects or triplet ones which depends on the choice of the four parameters.  In the following we carry out numerical analysis for these processes at a multi-TeV muon collider.

\section{Signatures in $\mu^-\mu^+ \to \tau^-\tau^+$ and $\mu^-\mu^+ \to \mu^\pm\mu^\pm + \tau^\mp \tau^\mp$}

\begin{figure}[!t]
	\centering
	\subfigure[\label{twobody1}]
	{\includegraphics[width=.4\textwidth]{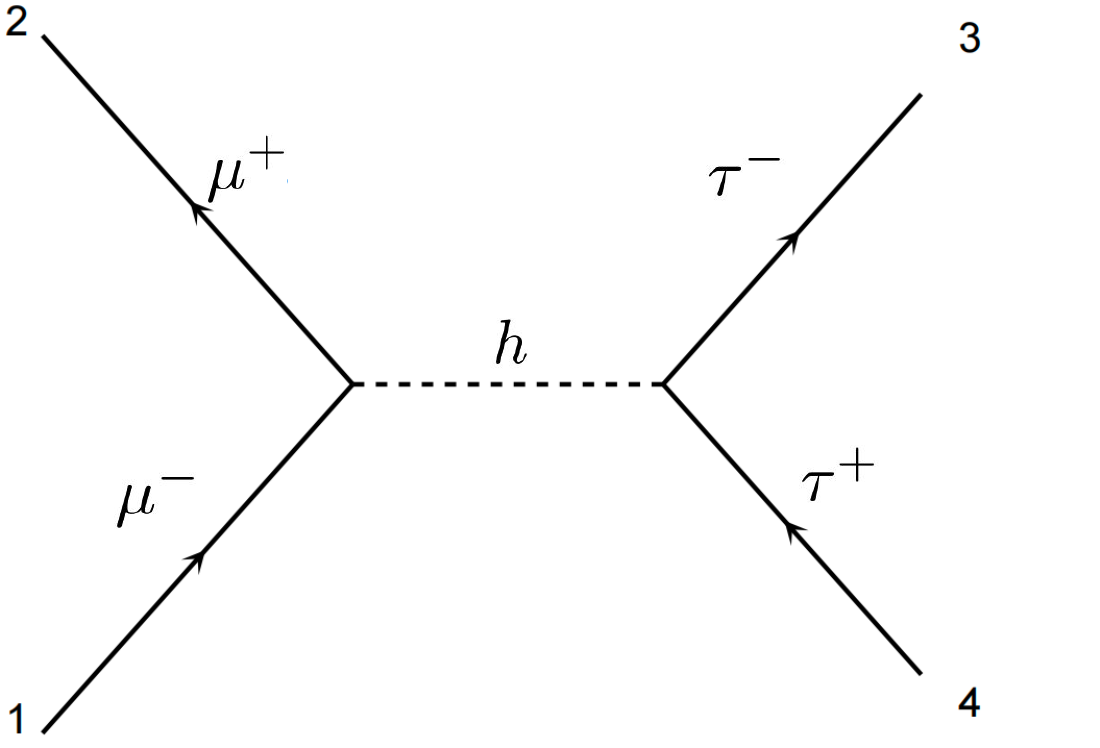}}
	\subfigure[\label{twobody2}]
	{\includegraphics[width=.4\textwidth]{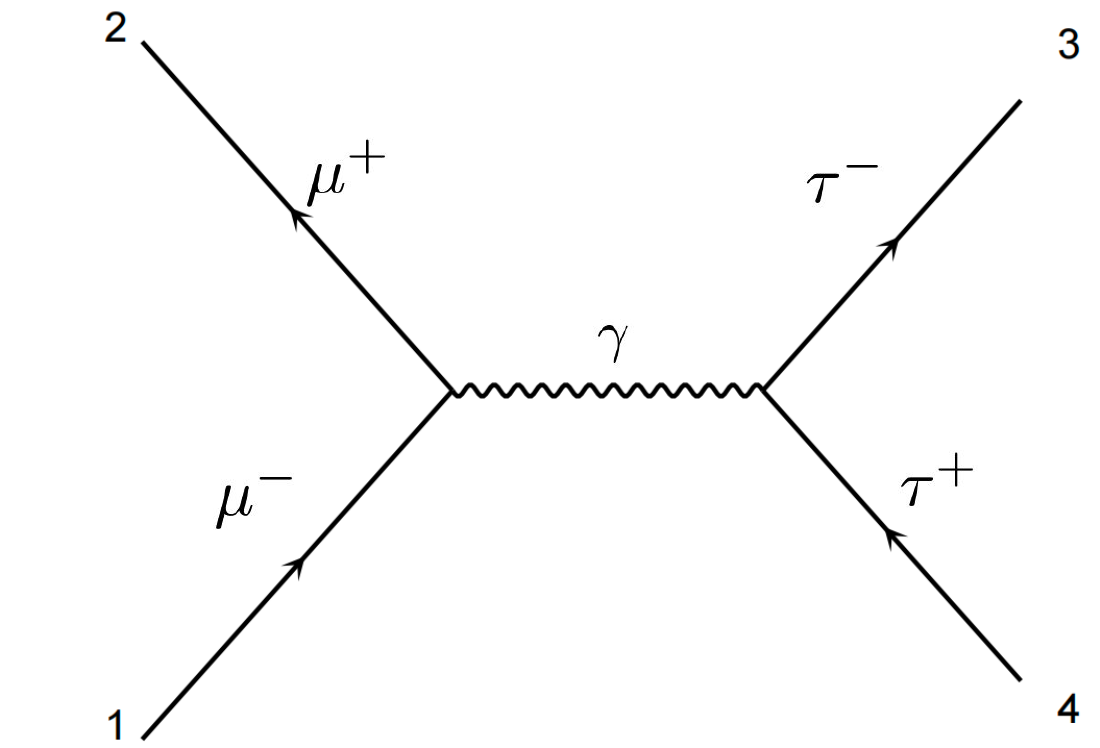}}
	\subfigure[\label{twobody3}]
	{\includegraphics[width=.4\textwidth]{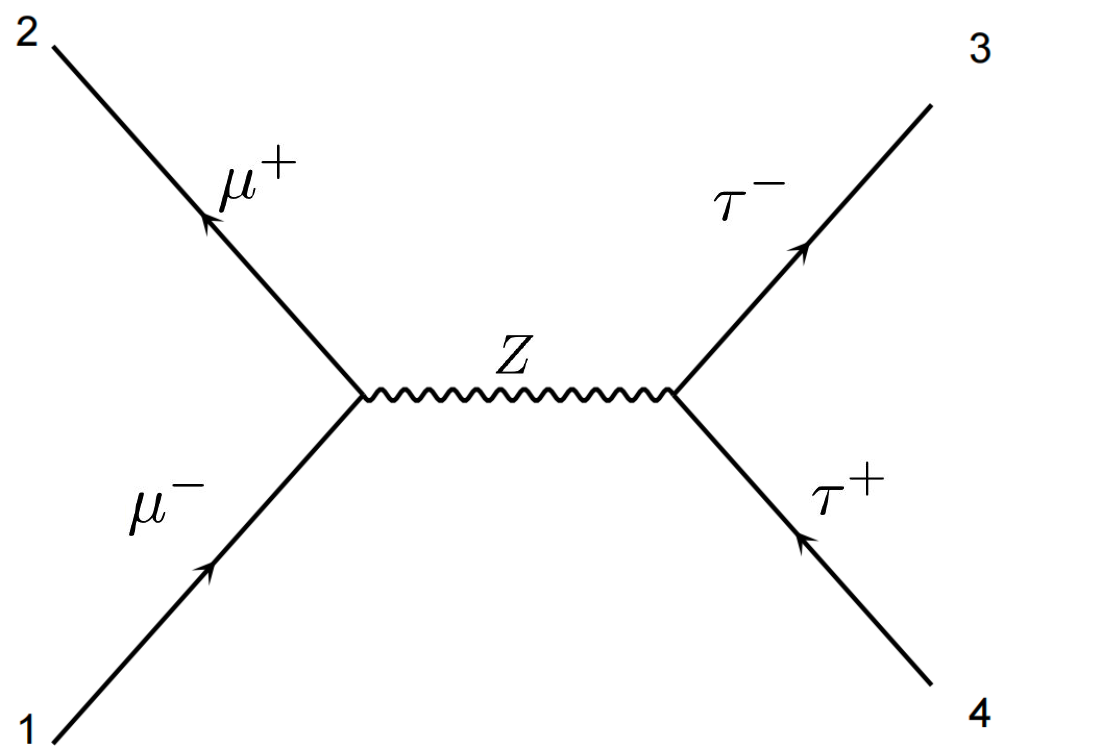}}
	\subfigure[\label{twobody4}]
	{\includegraphics[width=.4\textwidth]{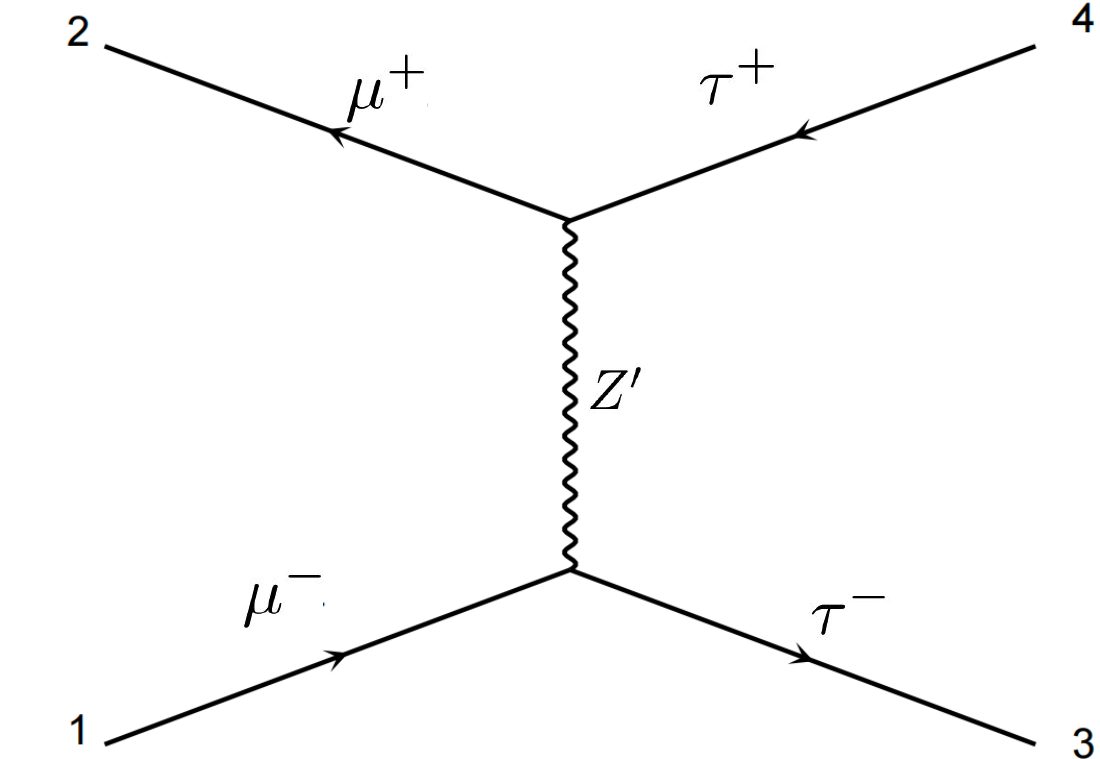}}
	\caption{The Feynman diagrams of flavor changing $\mu^-\mu^+ \to \tau^-\tau^+$ processes.  The SM s-channel contributions are shown in the first three figures, mediated by Higgs in Fig.(a), photon in Fig.(b) and Z boson in Fig.(c), respectively.  Fig.(d) means t-channel mediated by off-diagonal $Z'$ interaction. We do not show the contributions from the triplet effects, which can also contribute the process via s-channel. 
	}
	\label{twobodyfigure}
\end{figure}

\begin{figure}[!t]
	\centering
	\subfigure[\label{fourbody1}]
	{\includegraphics[width=.4\textwidth]{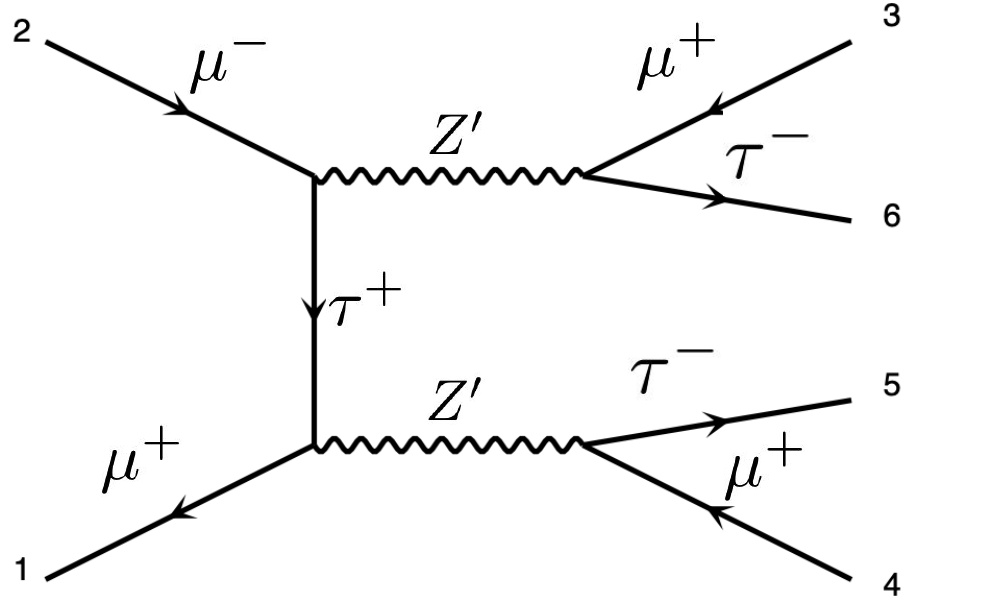}}
	\subfigure[\label{fourbody2}]
	{\includegraphics[width=.4\textwidth]{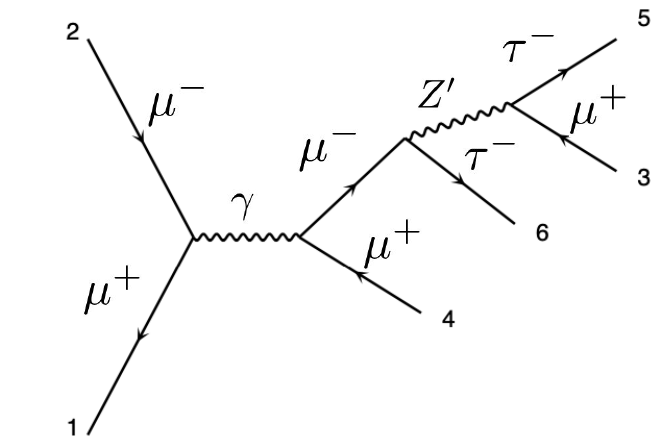}}
	\caption{The Feynman diagrams of flavor changing $\mu^-\mu^+ \to \mu^+\mu^+ + \tau^- \tau^-$ processes.  The diagrams are so much that we just show two examples. Here we only show the flavor changing $Z'$ effects. 
	}
	\label{fourbodyfigure}
\end{figure}

The maximal $\mu-\tau$ interaction of $Z'$ will affect $\mu^-\mu^+ \to \tau^- \tau^+$ via t-channel, which is different from the SM s-channel contributions  as shown in Fig.~\ref{twobodyfigure}.   Besides,  the  $Z'$ interaction can produce  distinctive signature $\mu^-\mu^+ \to \mu^\pm\mu^\pm + \tau^\mp \tau^\mp$  as shown in Fig.~\ref{fourbodyfigure}, which serves as a smoking gun for the model studied here. 

 For a multi-TeV muon collider,  the  luminosity scaling with energy quadratically~\cite{Delahaye:2019omf, Li:2023ksw} as 
	\begin{eqnarray}
	{\cal L} \geq	
 \left(\frac{\sqrt{s}}{10 \mbox{TeV}}\right)^2 \times 2\times 10^{35} \mbox{cm}^{-2} \mbox{s}^{-1} \;.
\end{eqnarray}
In particular, the benchmark choices of the collider energies and the corresponding integrated luminosities within 5 years is
	\begin{eqnarray}\label{muon}
	\sqrt{s}=	3 \;
 \; \mbox{TeV} \longrightarrow	{\cal L}= 1 
 \;ab^{-1} \;.
\end{eqnarray}

A very good identification of the leptons is a basic ingredient of many analyses at the colliders. In particular $\tau$-leptons, which are the most difficult leptons to identify, are expected to be produced by the decay of several interesting physic channels. Tau tagging is performed to identify jets likely to originate from a tau lepton, which is conducted by the hadronic decay mode of the taus. In the SM, tau decays hadronically with a probability of 65\%, producing a tau-jet mostly containing neutral and charged pions. In our case with a pair of taus in the final state, 42\% of the events will contain two tau-jets. The hadronic tau decays have low charged track multiplicity (one or three prongs) and a relevant fraction of the electromagnetic energy deposition due to photons coming from the decay of neutral pions. Moreover, when the momentum of the tau is large compared to its mass, the tau-jets will be highly collimated and produce localized energy deposits in the electromagnetic and hadronic calorimeters. These characteristics can be exploited to enhance the identication of hadronic tau decays. 
  At the muon collider,  the muon tagging efficiency is 100\% with $\eta<2.5$~\cite{Li:2023lin} and the $\tau$ tagging efficiency is 80\% with $p_{T} > 10$ GeV, as defined in Delphes cards~\cite{Frixione:2021zdp}.

To obtain simulation data for analysis, we implement interactions in the previous model 
through means of FeynRules~\cite{Alloul:2013bka}, we generated a Universal Feynman rules Output(UFO) model~\cite{Degrande:2011ua} for the model Lagrangian. Then fed the model into MadGraph5-aMC@NLO~\cite{Alwall:2011uj} for all simulations,  which are then fed to PYTHIA 8~\cite{Sjostrand:2014zea} for showering and hadronization, and DELPHES~\cite{deFavereau:2013fsa} for a fast detector simulation.
  

\subsection{ $\mu^+ \mu^- \to \tau^+ \tau^-$ analysis}

\begin{figure}[!t]
	\centering
	\includegraphics[width=0.4\textwidth]{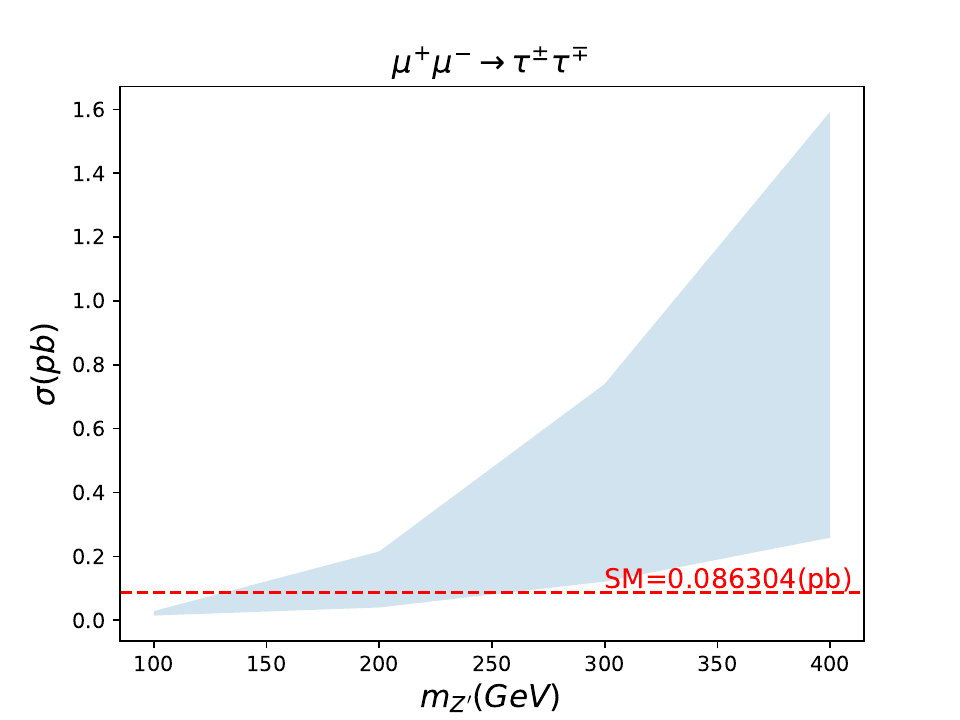}
  {\includegraphics[width=.4\textwidth]{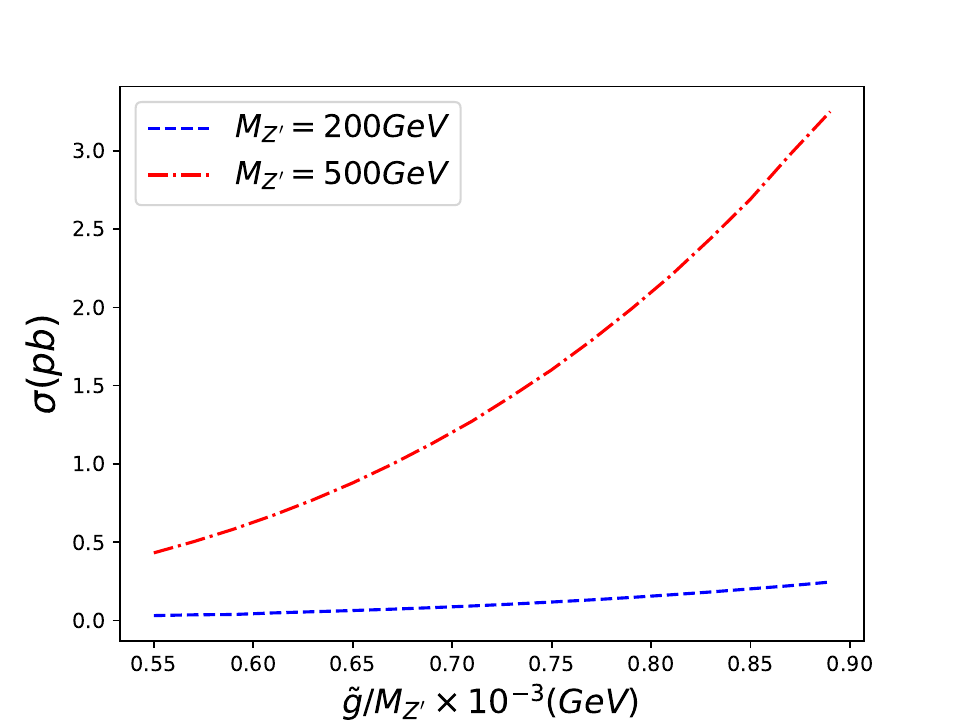}}
	\caption{The cross section of flavor changing $\mu^+ \mu^- \to \tau^+ \tau^-$ processes  for fixing  $m_\Delta=450 \mbox{GeV},\;  |Y_{22}|=0.117$.\ 
  The Left panel means the  ranges of the cross section with different $m_{Z'}$ in basic cuts. The blue regions show the allowed regions when varying the coupling constant $\tilde g$, and  the dashed red line means the SM background. The right panel shows the cross section with the ratio $\tilde g/m_{Z'}$ for two different $Z'$ mass cases with $P_T>250$ GeV.  
	}
	\label{flavorchanging}
\end{figure}

To extract signatures, we need to have a good understanding of the background. The SM backgrounds for $\mu^+ \mu^- \to \tau^+ \tau^-$ are shown in Table.~\ref{back}.   Here we use the following basic cuts~\cite{Li:2023lin}:
  (i) transverse momentum  $p_{T}>10$ GeV, 
  (ii) absolute pseudo-rapidity $|\eta|<2.5$,
  (iii) the separation  of the two leptons $\Delta R=\sqrt{(\Delta \eta)^2 +(\Delta \phi)^2}>0.4$.
  In addition to the s-channel scattering process with SM mediators $\gamma,\; Z,\;h$, there also exist other backgrounds from the blind features of the detector, 
   $\tau^{+}\tau^{-}\gamma$, $h/Z/\gamma (\to \tau^+ \tau^-)\;\nu\bar \nu$,  $W^{+}(\to \tau^+ \nu_\tau)W^{-}(\to \tau^- \bar \nu_\tau)$.  
  Therefore, the total backgrounds are 0.086304 pb as shown in  Fig.~\ref{flavorchanging}.
   This background value can be suppressed to 0.0144 pb by changing the cut $P_T > 250$ GeV as shown in Table.~\ref{back}.

   \begin{table}
  	\centering
  	\caption{   The  cross section of SM background $ \tau^\pm \tau^\mp$   at $\sqrt{s}=3$ TeV.}
  	\begin{tabular}{|c|c|c|c|c|c|c|c|c|}
  		\hline 
  		$\sigma$(pb)& $\tau^{+}\tau^{-}$   & $\tau^{+}\tau^{-}\gamma$ & $\gamma(\to \tau^+ \tau^-) \nu\bar\nu$ & $Z(\to \tau^+ \tau^-) \nu\bar\nu$ & $h(\to \tau^+ \tau^-) \nu\bar\nu$   & $W^{+}(\to \tau^+ \nu_\tau)W^{-}(\to \tau^- \bar \nu_\tau)$ &Total
  		\tabularnewline 
  		\hline  
  		1. Basic Cut &  0.008
  		& 	0.00162304 & 0.016896 &0.043552 & 0.0126848  & 0.0035424 &0.086304\\
  		\hline
  		2. $P_{T}> 250$ GeV &  0.007808
  		& 	0.00032 &  0.003584 & 0.00192 & 0.000064  & 0.000704 & 0.0144 \\
  		\hline
  	\end{tabular}  
  	\label{back}
  \end{table}

For the effects of the model under consideration,  we can obtain the corresponding cross section using model parameters depicted earlier.  
Based on the analysis in Ref.~\cite{Cheng:2021okr}, we use $\tilde g/m_{Z'}=(0.55, 0.89)\times 10^{-3} \mbox{GeV}^{-1}$  and $Y_{22}/m_\Delta=(0.26,1.42)\times 10^{-3} \mbox{GeV}^{-1}$ for resolving the muon $(g-2)_\mu$ anomaly while satisfying the other experimental constraints.  
The LHC search for a new $Z'$ gauge boson by the four muon (4$\mu$) final states, which excludes the coupling strength $\tilde g$ above 0.003-0.2 for $Z'$ mass ranging from 5 to 81 GeV at ATLAS~\cite{ATLAS:2023vxg} and $\tilde g$ above 0.004-0.3 for $Z'$ mass ranging from 5 to 70 GeV at CMS~\cite{CMS:2018yxg}. In fact, the above direct LHC constraints on $Z'$ from simple resonance searches like $pp \to Z' \to ll/jj$ are not applicable in our case, since the $Z'$ does not couple to quarks at the tree level. Moreover, the flavor-violating $Z'$ searches at the LHC have only focused on the $e\mu$ channel so far~\cite{ATLAS, CMS}. If assuming the constraints might also be used to analyze our case, we can choose $m_{Z'}>81$ GeV to evade the above bounds, which is actually the electroweak scale.  At the electroweak scale $U(1)_{L_\mu-L_\tau}$ $Z'$ has been shown to be allowed by experimental data~\cite{Heeck:2011wj}. Therefore, we focus on  $m_{Z'}\geq 100$ GeV.

In order to satisfy  the muon $(g-2)_\mu$ and other experimental constraints, we choose the triplet scalar parameters to  be~\cite{Cheng:2021okr} $m_\Delta=450 \mbox{GeV},\;  |Y_{22}|=0.117$. In this case, we find that the  triplet effects only lead to the cross section $\sigma=0.00944$ pb, whose contribution is so small  only  with around $1\%$ for the large $Z'$ case. 
We plot the cross section of flavor changing  $\mu^+ \mu^- \to \tau^+ \tau^-$ processes shown in Fig.~\ref{flavorchanging}.  
In this figure, the left panel is plotted for the cross section as a function of  $Z'$ mass and the right panel as a function of the ratio $\tilde g/m_{Z'}$. 
 The left panel shows  the allowed ranges of the cross section for different parameters $\tilde g$ and $m_{Z'}$ with the basic cuts. We found that the cross section highly depends on the $Z'$ mass and the gauge coupling $\tilde g$. And in the case of the small $Z'$  mass, the total cross section is smaller than the SM contribution so that the interference effects between SM and $Z'$ will reduce the SM contribution $\sigma=0.086304$ pb.  To further reduce the SM background, we impose the cut $P_T>250$ GeV as shown in Table.~\ref{back}.  Then we plot the right panel showing the cross section with the ratio $\tilde g/m_{Z'}$ for two different $Z'$ mass cases.

\begin{table}[!t]
	\centering
	\caption{  The cross section $ \tau^\pm \tau^\mp$  for $U(1)_{L_\mu-L_\tau}$ model with $Y=1$ triplet  at $\sqrt{s}=3$ TeV for fixing $m_\Delta=450$ GeV and $Y_{22}=0.117$.  }
	\begin{tabular}{|c|c|c|c|c|c|c|c|c|c|c|c|c|}
		\hline
		\multirow{2}{*}{$U(1)_{L_\mu-L_\tau}$ with triplet model} & \multicolumn{2}{|c|}{$m_{Z^{\prime}}=500$ GeV}  &
   \multicolumn{2}{|c|}{$m_{Z^{\prime}}=200$ GeV} &
   \multicolumn{2}{|c|}{$m_{Z^{\prime}}=100$ GeV}  \\		
             \cline{2-7}
		& $\tilde g =0.275$   & $\tilde g=0.445$  & $\tilde g=0.11$   & $\tilde g=0.178$  & $\tilde g=0.055$   & $\tilde g =0.089$  
        \tabularnewline 
  		\hline  
		cross section (pb) &  0.274
		 & 2.08	 & 0.017  &0.154 &0.0014 &0.01  \\
		\hline
  luminosity (fb$^{-1}$) with  $3\sigma$ 
  &0.034  &  0.004& 0.97&0.063 & 72.2335& 2.189  \\
		\hline
Events ( ${\cal L}=1 ab^{-1}$ )
 &274000 &2080000  & 17000 &154000  &1400 &10000   \\
		\hline
	\end{tabular}
	\label{twobodysectionXY}
\end{table}

  The influence of new physics is shown by the difference ratio factor  $(\sigma-\sigma_{SM})/\sigma_{SM}$, which is further translated  into a necessary luminosity  to discover a given scenario by defining a test statistic  $S/\sqrt{S+S_0}$. Here  $S = {\cal L} \times (\sigma-\sigma_{SM})$ means the new physics signal, and $S_0 ={\cal L} \times \sigma_{SM}$ means the background  ( ${\cal L}$ is the luminosity).
  Requiring $S/\sqrt{S_0}>3\; \mbox{or}\; 5$, we can assign a rough discovery luminosity to each  different case. The relevant information is shown in Table.~\ref{twobodysectionXY}.
 Based on the values, we can obtain the cross section  for $U(1)_{L_\mu-L_\tau}$ with Y=1 triplet model models.
Afterward, we can change different parameters to obtain the corresponding cross section and events. 
Note that the two values of $\tilde g$ for every fixed $m_{Z'}$ correspond to the lower and upper bounds, respectively. 
For example, for $m_{Z'}=100$ GeV case,  the lower bound $\tilde g=0.055$ can  generate  1400 events and the upper bound $g=0.089$ with 10000 events,  when the luminosity is $1ab^{-1}$ as required in Eq.~\ref{muon}.

 \begin{figure}[!t]
	\centering
	\includegraphics[width=0.4\textwidth]{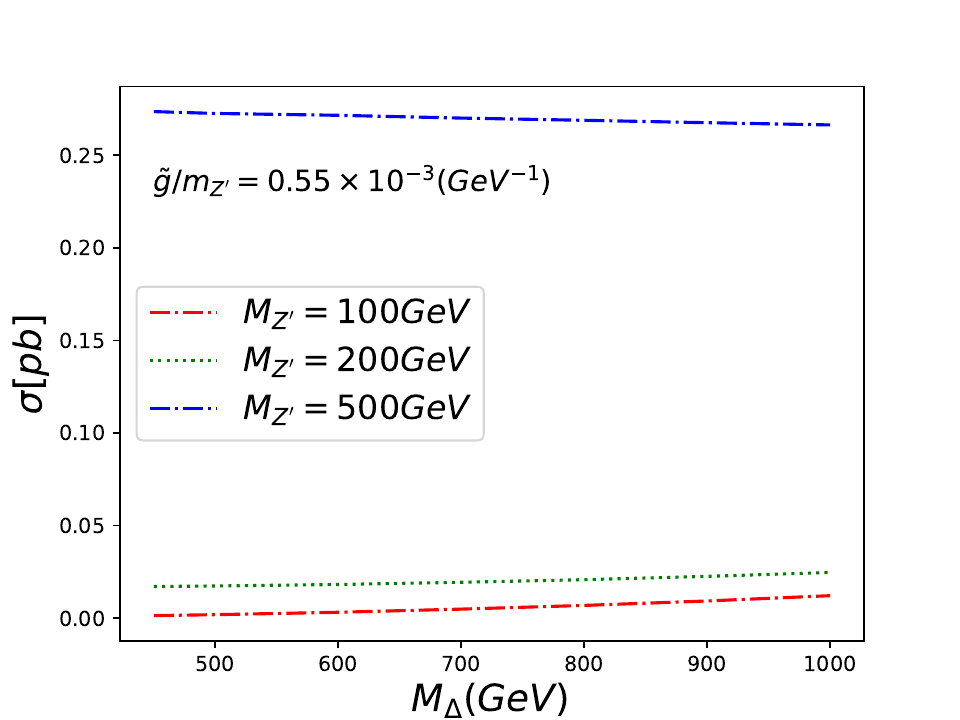}
	\caption{  The cross section of flavor changing $\mu^+ \mu^- \to \tau^+ \tau^-$ processes  with different $m_{\Delta}$ in basic cuts and  $P_{T}>250$ GeV. 
 Here we choose the lower bound $Y_{22}/m_\Delta=0.26\times 10^{-3} \mbox{GeV}^{-1}$.  The different color lines mean the different $Z'$ mass with fixed $\tilde g/m_{Z'}=0.55\times 10^{-3} \mbox{GeV}^{-1}$, 
 $m_{Z'}=100$ GeV in red,  $m_{Z'}=200$ GeV in green,  $m_{Z'}=500$ GeV in blue, respectively.
	}
	\label{triplet-m2}
\end{figure}

The above analysis is based on the choice of $m_\Delta=450 \mbox{GeV},\;  |Y_{22}|=0.117$. Actually, we also study the different contributions from triplet effects as shown in Fig.~\ref{triplet-m2}. 
We plot the cross section of flavor changing $\mu^+ \mu^- \to \tau^+ \tau^-$ processes  with different $m_{\Delta}$ in basic cuts and  $P_{T}>250$ GeV. Here we choose the lower bound $Y_{22}/m_\Delta=0.26\times 10^{-3} \mbox{GeV}^{-1}$ and fix $\tilde g/m_{Z'}=0.55\times 10^{-3} \mbox{GeV}^{-1}$ with three different $Z'$ masses. We found that the cross section varies slightly when increasing $\Delta$ mass.  
The corresponding luminosity and events are shown in Table.~\ref{twobodytriplet}. We found that the cross section is around $10^{-3}$ pb for the above two lower bounds, which is so smaller than the large $Z'$ mass effects.
This means that the choice of the  triplet parameters will not affect cross section to a large extent.
Therefore, we choose the previous parameters $m_\Delta=450 \mbox{GeV},\;  |Y_{22}|=0.117$ to conduct the analysis  in the following.

\begin{table}[!t]
	\centering
	\caption{  The cross section $ \tau^\pm \tau^\mp$  for $U(1)_{L_\mu-L_\tau}$ model  with $Y=1$ triplet  at $\sqrt{s}=3$ TeV for fixing $m_{Z^{\prime}}=100$ GeV and $\tilde g =0.055$.  }
	\begin{tabular}{|c|c|c|c|c|c|c|c|c|c|c|c|c|}
		\hline
		\multirow{2}{*}{$U(1)_{L_\mu-L_\tau}$ with triplet model} & \multicolumn{2}{|c|}{$m_{\Delta}=800$ GeV}  &
   \multicolumn{2}{|c|}{$m_{\Delta}=500$ GeV} &
   \multicolumn{2}{|c|}{$m_{\Delta}=450$ GeV}  \\		
             \cline{2-7}
		& $Y_{22} =0.208$   & $Y_{22} =1.136$  & $Y_{22} =0.13$   & $Y_{22} =0.71$  & $Y_{22} =0.117$   & $Y_{22}  =0.639$  
        \tabularnewline 
  		\hline  
		cross section (pb) &  0.0069
		 & 2.6156	 & 0.0019  &0.5180 &0.0014 &0.3569  \\
		\hline
  luminosity (fb$^{-1}$) with  $3\sigma$ 
  &4.00  &  0.0035& 39.287 &0.018 & 72.2335& 0.026   \\
		\hline
Events ( ${\cal L}=1 ab^{-1}$ )
 &6900 &2615600  & 1900 &518000  &1400 &356900   \\
		\hline
	\end{tabular}
	\label{twobodytriplet}
\end{table}

 \begin{figure}[!t]
	\centering
	\subfigure[\label{zpmasslumi}. The required luminosity for $3\sigma$ and $5\sigma$ discovery with $\tilde g / M_{Z^{\prime}}=0.55 \times 10^{-3} \mbox{GeV}^{-1}$ . ]
	{\includegraphics[width=.4\textwidth]{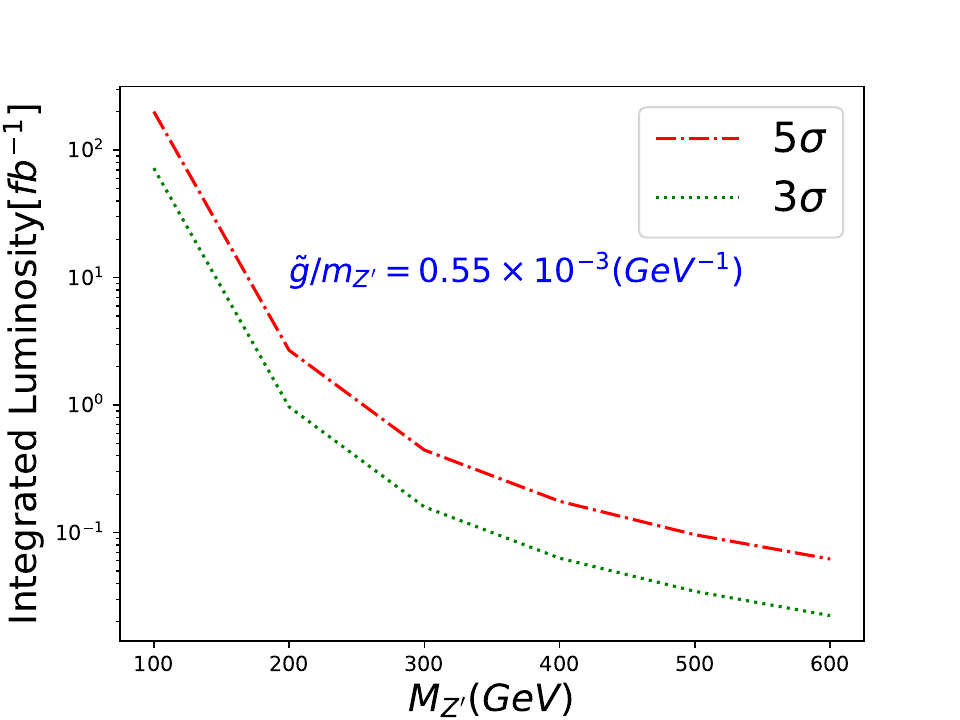}}
	\subfigure[\label{SU}. The required luminosity for $3\sigma$ and $5\sigma$ discovery with different $Z^{\prime}$ mass.]
	{\includegraphics[width=.4\textwidth]{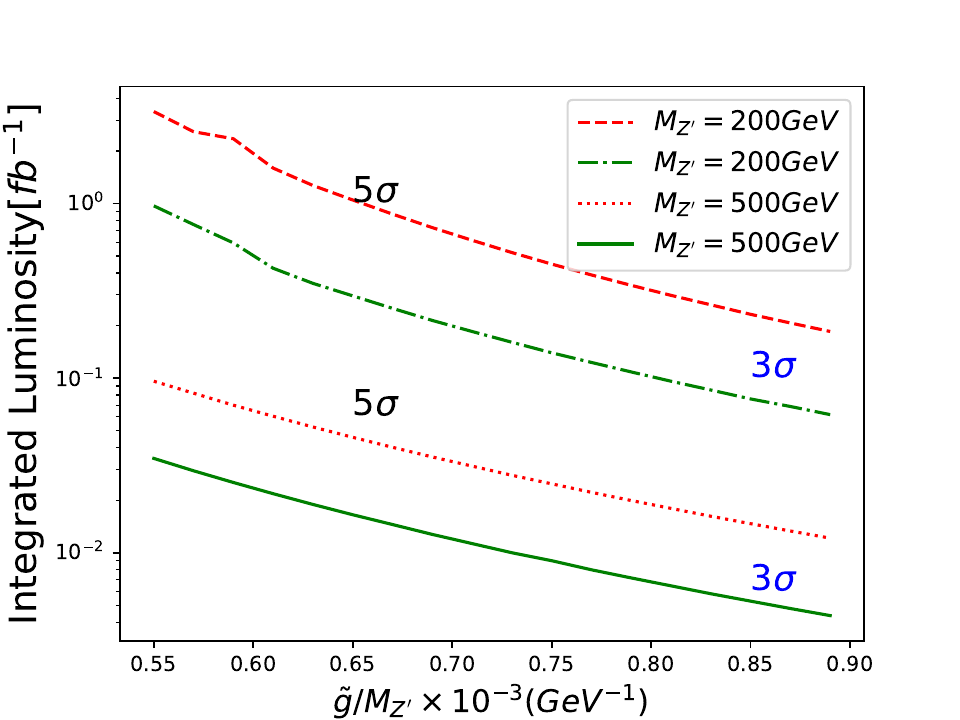}}
	\caption{ The required luminosity for different scenarios in the case of $\mu^{+}\mu^{-}\to\tau^{+}\tau^{-}$. The $3\sigma$ and $5\sigma$ significance are shown in green and red, respectively. }
	\label{zp}
\end{figure}

In order to further investigate the detection possibility of the flavoring changing process  $\mu^{+}\mu^{-}\to\tau^{+}\tau^{-}$ at the future muon collider, we present the required luminosity for $U(1)_{L_\mu-L_\tau}$ model with triplet scalar $Y=1$ as shown in Fig.~\ref{zp}.  Here we consider two different cases with   significance  $3\sigma$ and $5\sigma$.  In the left panel, we show the trend  of integrated luminosity with different $Z'$ mass when fixing the lower bound $\tilde g/m_{Z'}=0.55\times 10^{-3}$ GeV.
 We found that the luminosity will drop  rapidly with the original $\mathcal{O}$(100) $fb^{-1}$ into the final $\mathcal{O}$(0.02) $fb^{-1}$  when increasing $m_{Z'}$ from 100 GeV to 600 GeV.
In the right panel, we show the relation between the luminosity  and ratio    $\tilde g/m_{Z'}$.  Similarly, the luminosity shows the rapid fall trend when rising  $\tilde g/m_{Z'}$.  
 If increasing the ratio to the upper bound $\tilde g/m_{Z'}=0.89\times 10^{-3} \mbox{GeV}^{-1}$, the cross section will increase so that  the required luminosity will decrease with around an order of magnitude.

\subsection{$\mu^{-}\mu^{+} \rightarrow \mu^{\pm} \mu^{\pm} \tau^{\mp} \tau^{\mp}$ analysis}

Note that our model can  produce distinctive signature of the final states $\mu^\pm \mu^\pm \tau^\mp \tau^\mp$.   
This signal is very clean and effectively background-free. Although the tau reconstruction poses some practical challenges, the four lepton final states with same sign could provide a `smoking gun' signal for our scenario.
This unique signature has two different sources, $\mu^+\mu^-  \to h^*/\gamma^*/Z^*\to \mu^\pm \tau^\mp + (Z'\to \mu^\pm  \tau^\mp)$ and $\mu^+\mu^-  \to \Delta^{++}\Delta^{--} \to \mu^\pm \mu^\pm \tau^\mp \tau^\mp$. The two sources both contribute the cross section of same sign lepton pair final states. We should consider the effects of two sources simultaneously to identify the dominant contribution. 

We simulate the processes to estimate the sensitivity reach at $\sqrt{s} = 3$ TeV muon collider.  Due to the negligible SM background, we can impose the basic trigger cuts in the previous $\mu^+ \mu^- \to \tau^+ \tau^-$ case rather than changing $P_T$ additionally. Besides, we further impose the following cuts: 
 the leading lepton must satisfy the transverse momentum cut $p_T > 20$ GeV, while the sub-leading leptons are required to satisfy a milder cut $p_T > 15$ GeV.
These values are set to be as inclusive as possible for an optimistic analysis.

\begin{figure}[!t]
	\centering
	\includegraphics[width=0.4\textwidth]{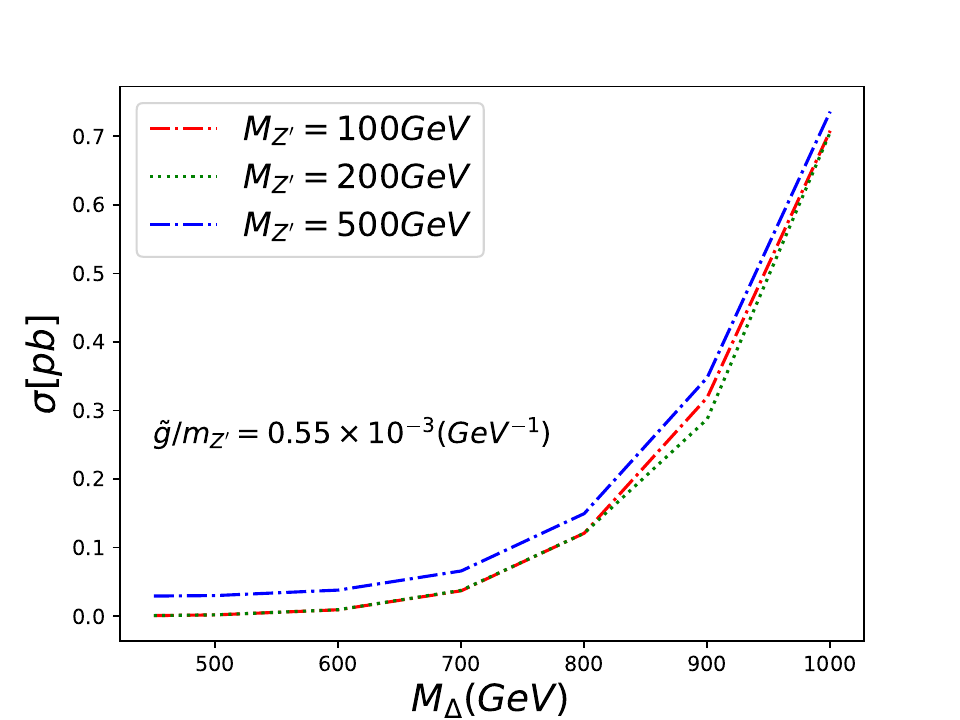}
	\caption{ The cross section of flavor changing $\mu^-\mu^+ \to \mu^\pm \mu^\pm + \tau^\mp\tau^\mp$ processes  with different $m_{\Delta}$ in basic cuts. 
		Here we choose the lower bound $Y_{22}/m_\Delta=0.26\times 10^{-3} \mbox{GeV}^{-1}$.  The different color lines mean the different $Z'$ mass with fixed $\tilde g/m_{Z'}=0.55\times 10^{-3} \mbox{GeV}^{-1}$, 
		$m_{Z'}=100$ GeV in red,  $m_{Z'}=200$ GeV in green,  $m_{Z'}=500$ GeV in blue, respectively.
	}
	\label{triplet-m4}
\end{figure}

We study the triplet effects on cross section of flavor changing $\mu^\pm \mu^\pm \tau^\mp \tau^\mp$ processes as shown in Fig.~\ref{triplet-m4}.   The  relation between the cross section and different $m_{\Delta}$ is plotted for fixing  the lower bound $Y_{22}/m_\Delta=0.26\times 10^{-3} \mbox{GeV}^{-1}$. We found that the triplet effects will rise when increasing the $\Delta$ mass. 
This is because that for fixed ratio $Y_{22}/m_\Delta$, increasing $m_\Delta$  leads to a bigger coupling $Y_{22}$. Also since the amplitude is proportional to $ |Y_{22}|^2/(s_1-m_{\Delta}^2)(s_2-m_{\Delta}^2)$ for $\mu^+\mu^-  \to \Delta^{++}\Delta^{--} \to \mu^\pm \mu^\pm \tau^\mp \tau^\mp$,  while increasing $m_{\Delta}^2$ makes  $(s_1-m_{\Delta}^2)(s_2-m_{\Delta}^2)$ smaller in the region of kinematics, therefore it results in a rise for the cross section as shown in Fig.~\ref{triplet-m4}.

This means that  the previous triplet parameters $m_\Delta=450 \mbox{GeV},\;  |Y_{22}|=0.117$ will give comparatively small contribution. Next we will choose these two values to study the feasible detection sensitivity. If further increasing $m_\Delta$, the cross section of $\mu^\pm \mu^\pm \tau^\mp \tau^\mp$ processes will improve several times, which is more feasible to be detected in the future muon collider.

Since the SM background is negligible for the doubly same sign dilepton pairs $\mu^\pm \mu^\pm \tau^\mp \tau^\mp$, we can simply estimate the signal sensitivity as $N=S/\sqrt{(S+S_0)} \approx \sqrt{{\cal L} \sigma_{signal}} $, where ${\cal L}$ is the integrated luminosity and $\sigma_{signal}$ is the signal cross section, as obtained from our detector simulation.  The corresponding cross sections and luminosity are shown in Table.~\ref{fourbodysectionXY}. 
Here we try to choose many different triplet parameters to achieve the aim of suppressing the triplet effects. According to our calculation, we found that when choosing $Y_{22}=0.117$  and $m_{\Delta}=450$ GeV, the triplet contribution is around $0.00075$ pb. 
 Compared to the values in Table.~\ref{fourbodysectionXY},  the triplet effects will weaken gradually along with the increase of $Z'$ mass, which makes $Z'$ effects more dominant.  For example $m_{Z'}=500$ GeV, the triplet effect is much smaller to be neglected  only with 2\% ratio proportion compared to $Z'$ contribution.

 \begin{table}[!t]
 	\centering
 	\caption{  The cross section $ \mu^\pm \mu^\pm \tau^\mp \tau^\mp$  for $U(1)_{L_\mu-L_\tau}$ model with $Y=1$ triplet  at $\sqrt{s}=3$ TeV for fixing $m_\Delta=450$ GeV and $Y_{22}=0.117$.  }
 	\begin{tabular}{|c|c|c|c|c|c|c|c|c|c|c|c|c|}
 		\hline
 		\multirow{2}{*}{$U(1)_{L_\mu-L_\tau}$ with triplet model} & \multicolumn{2}{|c|}{$m_{Z^{\prime}}=500$ GeV}  &
 		\multicolumn{2}{|c|}{$m_{Z^{\prime}}=200$ GeV} &
 		\multicolumn{2}{|c|}{$m_{Z^{\prime}}=100$ GeV}  \\		
 		\cline{2-7}
 		& $\tilde g =0.275$   & $\tilde g=0.445$  & $\tilde g=0.11$   & $\tilde g=0.178$  & $\tilde g=0.055$   & $\tilde g =0.089$  
 		\tabularnewline 
 		\hline  
 		cross section (pb) & 
 		0.029& 0.863	 &   0.00083&0.00134&	
 		0.00076  &  	
 		0.00079 \\
 		\hline
 		luminosity (fb$^{-1}$) with  $3\sigma$ 
 		& 0.31 &0.010  & 10.843 & 6.716& 11.842&  11.392  \\
 		\hline
 		Events ( ${\cal L}=1 ab^{-1}$ )
 		&29000 &863000  &830  & 1340 &760 & 790  \\
 		\hline
 	\end{tabular}
 	\label{fourbodysectionXY}
 \end{table}

\begin{figure}[!t]
	\centering
	\subfigure[\label{zpmasslumi}. The required luminosity for $3\sigma$ and $5\sigma$ discovery with $\tilde g / M_{Z^{\prime}}=0.55 \times 10^{-3} \mbox{GeV}^{-1}$ . ]
	{\includegraphics[width=.4\textwidth]{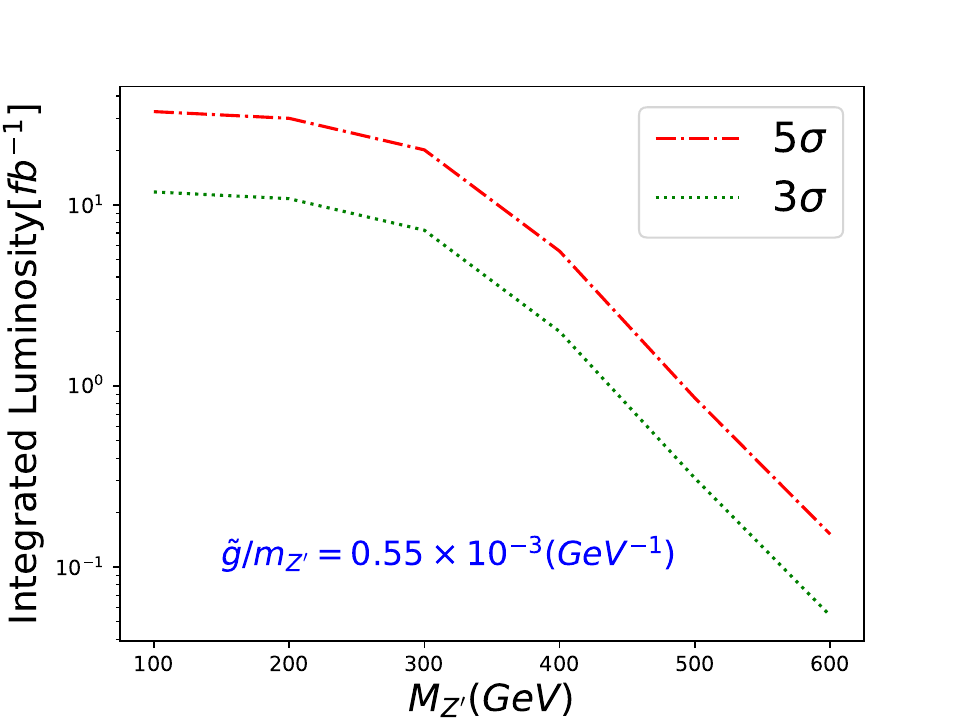}}
	\subfigure[\label{SU}. The required luminosity for $3\sigma$ and $5\sigma$ discovery with different $Z^{\prime}$ mass.]
	{\includegraphics[width=.4\textwidth]{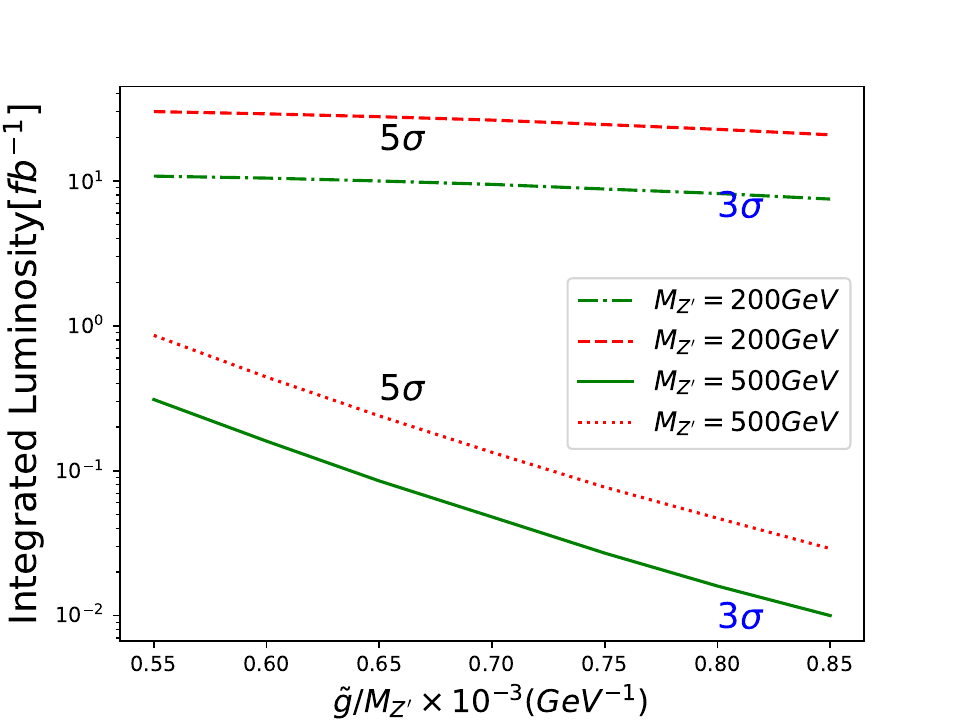}}
	\caption{ The required luminosity for different scenarios in the case of $\mu^-\mu^+ \to \mu^\pm \mu^\pm + \tau^\mp\tau^\mp$ with baisc cuts. The $3\sigma$ and $5\sigma$ significance are shown in green and red, respectively. }
	\label{fourzp}
\end{figure}

Similarly, we obtain the required luminosity for  $\mu^{+}\mu^{-} \to \mu^\pm \mu^\pm + \tau^\mp\tau^\mp$ to further investigate  detection possibility as shown in Fig.~\ref{fourzp}.    We analyze  the required luminosity with  significance  $3\sigma$ and $5\sigma$ in different cases  at the muon collider,  different $Z'$ mass with the fixed lower bound $\tilde g/m_{Z'}=0.55\times 10^{-3}$ GeV in the left panel and the ratio $\tilde g/m_{Z'}$ in the right panel.  We found that the required luminosity will tend to decrease.  In the left panel,  the required luminosity will gradually tend to abrupt when increasing $m_{Z'}$.
 In the right panel,  the luminosity shows the fall trend when rising  $\tilde g/m_{Z'}$ for two kinds of different $Z'$ masses. 
 If increasing the ratio into the upper bound  $\tilde g/m_{Z'}=0.89\times 10^{-3}$ GeV, the required luminosity will decrease with  different degrees depending on the $Z'$ mass.

Therefore, we find that the smoking gun signature of doubly same sign $\mu^\pm \mu^\pm + \tau^\mp\tau^\mp$ pairs production can have a 5$\sigma$ sensitivity, if required to solve the muon g-2 anomaly, at a muon collider of a 3 TeV with $\mathcal{O}$(fb) luminosity.
If further changing the triplet parameters to enhance the effects as shown in Fig.~\ref{triplet-m4}, the cross section increases rapidly so that the discovery potential will become more obvious.

\section{Conclusion}

 We have studied in detail the  maximal $\mu-\tau$ interaction of a $Z'$  in $U(1)_{L_\mu-L_\tau}$ model at a muon collider. 
The  maximal  $Z'$ off-diagonal mixing, $(\bar \mu \gamma^\mu \tau + \bar \tau \gamma^\mu \mu) Z'_\mu$, will escape  other constraints with  $Z'$ mass to be lower than a few hundred MeV while addressing the muon g-2 anomaly. In addition, a  $Z'$  with a large  mass  can result in  very distinctive signatures, such as muon collider t-channel production of $\mu^-\mu^+ \to \tau^- \tau^+$ pair, and  doubly same sign muon and tau pairs production, $\mu^- \mu^+ \to \mu^\pm\mu^\pm \tau^\mp \tau^\mp$.  
With a muon collider of 3 TeV with $\mathcal{O}$($fb^{-1}$) luminosity, we find that within the muon g-2 anomaly constrained parameter spaces for the ratio of $Z'$ coupling and mass, and triplet Higgs contributions, 
for the $\mu^-\mu^+ \to \tau^- \tau^+$ case, the t-channel pair production can be easily distinguished at more than 5$\sigma$  level from the s-channel production as that predicted in the standard model. 
For the $\mu^- \mu^+ \to \mu^\pm\mu^\pm \tau^\mp \tau^\mp$ case, it can serve as the smoking gun signature for our model which can be discovered at 5$\sigma$ level.

\begin{acknowledgments}
	
	This work was supported in part by Key Laboratory for Particle Physics, Astrophysics and Cosmology, Ministry of Education, and Shanghai Key Laboratory for Particle Physics and Cosmology (Grant No. 15DZ2272100),  in part by the NSFC (Grant Nos. 11735010, 11975149, and 12090064) and partially supported by the Fundamental Research Funds for the Central Universities. XGH was supported in part by the MOST (Grant No. MOST 106-2112-M-002-003-MY3).  
\end{acknowledgments}


\begin{thebibliography}{99}
	
	\bibitem{Muong-2:2021ojo}
	B.~Abi \textit{et al.} [Muon g-2],
	Phys. Rev. Lett. \textbf{126} (2021) no.14, 141801
	doi:10.1103/PhysRevLett.126.141801
	[arXiv:2104.03281 [hep-ex]].
	
	
	
	
	\bibitem{Baek:2001kca}
	S.~Baek, N.~G.~Deshpande, X.~G.~He and P.~Ko,
	Phys. Rev. D \textbf{64} (2001), 055006
	[arXiv:hep-ph/0104141 [hep-ph]].
	
	
	\bibitem{Ma:2001md}
	E.~Ma, D.~P.~Roy and S.~Roy,
	Phys. Lett. B \textbf{525} (2002), 101-106
	[arXiv:hep-ph/0110146 [hep-ph]].
	
	
	\bibitem{Gninenko:2001hx}
	S.~N.~Gninenko and N.~V.~Krasnikov,
	Phys. Lett. B \textbf{513} (2001), 119
	[arXiv:hep-ph/0102222 [hep-ph]].
	
	\bibitem{Pospelov:2008zw}
	M.~Pospelov,
	Phys. Rev. D \textbf{80} (2009), 095002
	[arXiv:0811.1030 [hep-ph]].
	
	\bibitem{Heeck:2011wj}
	J.~Heeck and W.~Rodejohann,
	Phys. Rev. D \textbf{84} (2011), 075007
	[arXiv:1107.5238 [hep-ph]].
	
	\bibitem{Harigaya:2013twa}
	K.~Harigaya, T.~Igari, M.~M.~Nojiri, M.~Takeuchi and K.~Tobe,
	JHEP \textbf{03} (2014), 105
	[arXiv:1311.0870 [hep-ph]].
	
	
	
	\bibitem{Altmannshofer:2014pba}
	W.~Altmannshofer, S.~Gori, M.~Pospelov and I.~Yavin,
	Phys. Rev. Lett. \textbf{113} (2014), 091801
	[arXiv:1406.2332 [hep-ph]].
	
	
	\bibitem{Altmannshofer:2016oaq}
	W.~Altmannshofer, M.~Carena and A.~Crivellin,
	Phys. Rev. D \textbf{94} (2016) no.9, 095026
	[arXiv:1604.08221 [hep-ph]].
	
	
	\bibitem{CCFR:1991lpl}
	S.~R.~Mishra \textit{et al.} [CCFR],
	Phys. Rev. Lett. \textbf{66} (1991), 3117-3120
	
	\bibitem{CHARM-II:1990dvf}
	D.~Geiregat \textit{et al.} [CHARM-II],
	Phys. Lett. B \textbf{245} (1990), 271-275
	
	\bibitem{NuTeV:1999wlw}
	T.~Adams \textit{et al.} [NuTeV],
	Phys. Rev. D \textbf{61} (2000), 092001
	[arXiv:hep-ex/9909041 [hep-ex]].
	
	
	
	
	\bibitem{Cen:2021ryk}
	J.~Y.~Cen, Y.~Cheng, X.~G.~He and J.~Sun,
	Nucl. Phys. B \textbf{978} (2022), 115762
	[arXiv:2104.05006 [hep-ph]].
	
	
	\bibitem{Cheng:2021okr}
	Y.~Cheng, X.~G.~He and J.~Sun,
	Phys. Lett. B \textbf{827} (2022), 136989
	[arXiv:2112.09920 [hep-ph]].

 \bibitem{lmu-ltau1}
X.~G.~He, G.~C.~Joshi, H.~Lew and R.~R.~Volkas,
Phys. Rev. D \textbf{43} (1991), R22.

\bibitem{lmu-ltau2}
X.~G.~He, G.~C.~Joshi, H.~Lew and R.~R.~Volkas,
Phys. Rev. D \textbf{44}, 2118-2132 (1991)
	
	\bibitem{Foot:1994vd}
	R.~Foot, X.~G.~He, H.~Lew and R.~R.~Volkas,
	Phys. Rev. D \textbf{50} (1994), 4571-4580.
	
	

	
	
	
	
	










\bibitem{Lazarides:1980nt}
G.~Lazarides, Q.~Shafi and C.~Wetterich,
Nucl. Phys. B \textbf{181} (1981), 287-300.


\bibitem{Mohapatra:1980yp}
R.~N.~Mohapatra and G.~Senjanovic,
Phys. Rev. D \textbf{23} (1981), 165.

\bibitem{Konetschny:1977bn}
W.~Konetschny and W.~Kummer,
Phys. Lett. B \textbf{70} (1977), 433-435.

\bibitem{Cheng:1980qt}
Cheng,T.P. and Li,L.F.,
Phys. Rev. D \textbf{22}, 2860 (1980).

\bibitem{Magg:1980ut}
M.~Magg and C.~Wetterich,
Phys. Lett. B \textbf{94} (1980), 61-64.

\bibitem{Schechter:1980gr}
Schechter,J. and Valle,J.W.F.,
Phys. Rev. D \textbf{22}, 2227 (1980).

\bibitem{Ashanujjaman:2021txz}
S.~Ashanujjaman and K.~Ghosh,
JHEP \textbf{03} (2022), 195
[arXiv:2108.10952 [hep-ph]].

\bibitem{Delahaye:2019omf}
J.~P.~Delahaye, M.~Diemoz, K.~Long, B.~Mansouli\'e, N.~Pastrone, L.~Rivkin, D.~Schulte, A.~Skrinsky and A.~Wulzer,
[arXiv:1901.06150 [physics.acc-ph]].


\bibitem{Li:2023ksw}
T.~Li, C.~Y.~Yao and M.~Yuan,
JHEP \textbf{03} (2023), 137
[arXiv:2301.07274 [hep-ph]].




\bibitem{Li:2023lin}
J.~Li, W.~Wang, X.~Cai, C.~Yang, M.~Lu, Z.~You, S.~Qian and Q.~Li,
[arXiv:2302.02203 [hep-ph]].


\bibitem{Frixione:2021zdp}
S.~Frixione, O.~Mattelaer, M.~Zaro and X.~Zhao,
[arXiv:2108.10261 [hep-ph]].





\bibitem{Alloul:2013bka}
A.~Alloul, N.~D.~Christensen, C.~Degrande, C.~Duhr and B.~Fuks,
Comput. Phys. Commun. \textbf{185} (2014), 2250-2300
[arXiv:1310.1921 [hep-ph]].


\bibitem{Degrande:2011ua}
C.~Degrande, C.~Duhr, B.~Fuks, D.~Grellscheid, O.~Mattelaer and T.~Reiter,
Comput. Phys. Commun. \textbf{183} (2012), 1201-1214
[arXiv:1108.2040 [hep-ph]].


\bibitem{Alwall:2011uj}
J.~Alwall, M.~Herquet, F.~Maltoni, O.~Mattelaer and T.~Stelzer,
JHEP \textbf{06} (2011), 128
[arXiv:1106.0522 [hep-ph]].





\bibitem{Sjostrand:2014zea}
T.~Sj\"ostrand, S.~Ask, J.~R.~Christiansen, R.~Corke, N.~Desai, P.~Ilten, S.~Mrenna, S.~Prestel, C.~O.~Rasmussen and P.~Z.~Skands,
Comput. Phys. Commun. \textbf{191} (2015), 159-177
[arXiv:1410.3012 [hep-ph]].

\bibitem{deFavereau:2013fsa}
J.~de Favereau \textit{et al.} [DELPHES 3],
JHEP \textbf{02} (2014), 057
[arXiv:1307.6346 [hep-ex]].









\bibitem{ATLAS:2023vxg}
 [ATLAS],
[arXiv:2301.09342 [hep-ex]].

\bibitem{CMS:2018yxg}
A.~M.~Sirunyan \textit{et al.} [CMS],
Phys. Lett. B \textbf{792} (2019), 345-368
[arXiv:1808.03684 [hep-ex]].



\bibitem{ATLAS}
The ATLAS collaboration, ATLAS-CONF-2015-072.

\bibitem{CMS}
CMS Collaboration, CMS-PAS-EXO-16-001.









\end{thebibliography}
\end{document}